\RequirePackage{fix-cm}
\documentclass[smallextended]{svjour3}       
\smartqed  
\usepackage{graphicx}
\usepackage{mathptmx}      
\usepackage{amssymb}
\usepackage{amsfonts}
\usepackage{amsmath}
\usepackage{amstext}
\usepackage{braket}
\usepackage{graphicx}
\usepackage{bm}
\usepackage{soul}
\usepackage{xcolor}
\usepackage{marginnote}
\usepackage[normalem]{ulem}
\usepackage{enumitem}
\usepackage{empheq}
\usepackage{bbold}
\usepackage{dsfont}
\usepackage{xcolor}
\providecommand{\U}[1]{\protect\rule{.1in}{.1in}}

 \journalname{Journal of Statistical Physics}
\begin{document}

\title{Nonequilibrium oscillations, probability
angular momentum, and the climate system
}

\titlerunning{Nonequilibrium oscillations}        

\author{Jeffrey B. Weiss        \and
             Baylor Fox-Kemper   \and
             Dibyendu Mandal      \and
             Arin D. Nelson            \and
            \\ R. K. P. Zia
}


\institute{Jeffrey B. Weiss \at
              Department of Atmospheric and Oceanic Sciences, University of Colorado, Boulder, CO 80309 \\
              \email{Jeffrey.Weiss@colorado.edu}\\
           \and
           Baylor Fox-Kemper \at
           Department of Earth, Environmental, and Planetary Sciences, Brown University, Providence, RI 02912 \\
           \email{baylor@brown.edu}\\
          \and
          Dibyendu Mandal \at
          Department of Physics, University of California, Berkeley, California 94720, USA \\
           \email{dibyendu.mandal@gmail.com}\\
          \and
          Arin D. Nelson \at
          Department of Earth and Environmental Sciences, University
          of Michigan, Ann Arbor, MI 48109 \\ 
          Tel.: +1-906-282-4466\\
          \email{dr.adnelson@gmail.com}\\
         \and
         R. K. P. Zia \at
         Department of Physics \& Astronomy, University of North Carolina at Asheville, Asheville, NC 28804, USA \& Center for Soft Matter and Biological Physics, Department of Physics, Virginia Polytechnic Institute and State University, Blacksburg, VA 24061, USA  \\
           \email{rkpzia@vt.edu}\\
}

\date{Submitted March 17, 2019, accepted September 23, 2019}

\maketitle

\begin{abstract}
Though the Boltzmann-Gibbs framework of equilibrium statistical mechanics 
has been successful in many arenas, it is clearly inadequate for describing 
many interesting natural phenomena driven far from equilibrium. The simplest step towards that goal is a better understanding of
nonequilibrium \textit{steady-states} (NESS). Here we focus on 
one of the distinctive features of NESS -- persistent probability currents --
and their manifestations in our climate system. We consider the natural variability
of the steady-state climate system, which can be approximated as a 
NESS. These currents must form closed loops, 
which are odd under time reversal, providing the crucial difference between
systems in thermal equilibrium and NESS. Seeking manifestations of such 
current loops leads us naturally to the notion of  
\textquotedblleft probability angular momentum\textquotedblright\ 
and oscillations in the space of observables. Specifically, we will
relate this concept to the \textit{asymmetric} part of certain 
time-dependent correlation functions. Applying this approach,
we propose that these current loops give rise to preferred spatio-temporal 
patterns of natural climate variability that take the form of climate 
oscillations such as the El-Ni\~{n}o Southern Oscillation (ENSO) and the 
Madden-Julien Oscillation (MJO). In the space of climate indices, we observe
persistent currents and define a new diagnostic for these currents: the
probability angular momentum ($\mathcal{L}$). Using the observed climatic time
series of ENSO and MJO, we compute both the averages and the distributions
of $\mathcal{L}$. These results are in good agreement with the
analysis from a linear Gaussian model. We propose that, in addition to 
being a new quantification of climate oscillations across models and 
observations, the probability angular momentum provides a meaningful 
characterization for all statistical systems in NESS.
\end{abstract}

\section{Introduction}

Nearly all the interesting phenomena around us emerge from tractable
interactions between simple constituents, e.g., electromagnetism and atoms.
However, understanding how emergent phenomena \cite{Schmidt2007} arise - the goal of
statistical mechanics - is extremely challenging. For systems in
thermal equilibrium, Boltzmann and Gibbs provided a highly successful
framework, while linear response theory is adequate for describing systems 
\textit{near} equilibrium, see e.g., \cite{ReichlPrigogine80}. Yet, most
fascinating phenomena in nature are associated with systems driven far from
equilibrium \cite{Gallavotti2014}, e.g., all life forms, socio-political structures, and the 
climate system. In particular, such systems would either not exist or be vastly different under conditions of
thermal equilibrium, i.e., when they are totally isolated or allowed to
exchange energy (or particles, or information \cite{ParrondoEtAl2015}) with
just one reservoir. Despite much progress on fluctuation theorems and the
\textquotedblleft nonequilibrium counterpart\textquotedblright\ of the free
energy in recent years (see, e.g., Ref. \cite{US12}), an overarching
framework for far-from-equilibrium systems remains elusive. Often, to study
such interesting systems, we rely on models with a few (macroscopic) degrees
of freedom, evolving as nonequilibrium stochastic processes. One frequently
used approach involves master or Fokker-Planck equations for the probability distribution,
with time-independent rates. While analyzing the full time dependence is
generally beyond our reach, we can take initial steps, by studying the
associated stationary states (which are guaranteed to exist). If these rates
obey detailed balance, the stationary distribution can be easily found and 
the system can be treated as if it is in thermal equilibrium 
\cite{ReichlPrigogine80}.
On the other hand, if the rates \textit{violate} detailed balance, then even
finding the stationary distribution is highly non-trivial in general 
\cite{Hill66}. Specifically, such detailed balance-violating systems settle
into nonequilibrium steady-states (NESS), and understanding their properties
(e.g., fluctuations and correlations) is quite challenging. In particular,
unlike systems in thermal equilibrium, there are persistent probability
currents that remain in the infinite time
  mean \cite{ZS07}, which form closed loops and characterize
underlying 
rotations in configuration space. Studying the observable
consequences of such steady current loops is surely a valuable endeavor,
 and is likely to lead to fruitful insights for all NESS. Here we focus one such
observable - the probability angular momentum, in analogy with the familiar
angular momenta associated with fluid current loops (e.g., 
\cite{SuzukiFox-Kemper16a}). As shown below, this quantity is intimately related
to fluctuations and temporal correlations in the NESS. Introduced recently
in other contexts \cite{SZ14,MMZ16,MMZ17,UGa16}, it will be considered here in the
context of the Earth's climate system.

The climate system is forced by incoming short-wave solar radiation and it is damped by 
outgoing infrared radiation
emitted to space, with a distribution of net radiative forcing segregated by latitude.
As a result, the climate system is approximately in a NESS \cite{Lorenz1955,PeixotoOort,PauluisHeld2002,Lucarini2009,LucariniEtAl2014,Laliberte2015}.
While this approximation is
violated by non-steady forcings such as solar variability, seasons and
Milankovitch cycles in the Earth's orbit, intermittent volcanic eruptions,
and anthropogenic greenhouse forcing, much remains to be learned about the
steady-state climate. In this study, we will ignore non-steady forcings. 

The climate system is known to exhibit many self-organized, irregular,
spatio-temporal patterns, typically referred to as oscillations. These
patterns include the El-Ni\~{n}o Southern Oscillation (ENSO) 
\cite{Wang2018,WangEtAl2017,MeinenMcPhaden00}, 
the Madden-Julien Oscillation (MJO) \cite{Zhang05}, 
the Pacific Decadal
Oscillation (PDO) \cite{NewmanAlexander16}, 
and the Atlantic
Multidecadal Oscillation (AMO) \cite{KnightFolland06}. 
It should be emphasized that these \textquotedblleft 
oscillations\textquotedblright\ are not single-frequency constant amplitude
sinusoidal fluctuations or necessarily wavelike phenomena: their frequency
distributions and amplitude variations are broad and important, but they are
narrow enough that each one is an empirically recognized coherent
spatio-temporal pattern of natural variability. ENSO and the MJO are
emergent phenomena that result from a complex \textquotedblleft
organization\textquotedblright\ of dynamical processes including tropical convection,
the velocities and temperatures of the atmosphere and ocean,
and large-scale oceanic and atmospheric waves \cite%
{NealeRichter08,KiladisWheeler09}. As such, they are unlikely to obey
detailed balance or any particular time-reversal symmetry. Nonetheless, ENSO
and the MJO are the dominant modes of equatorial interannual and tropical
intraseasonal variability, respectively. So, ENSO and the MJO both emerge
from a multiplicity of mechanisms, and dominate other variability in their
regions and timescales. These climate oscillations are seen as
fluctuations about the time mean climate state and we interpret the 
specific spatio-temporal character of
the oscillations as the physical-space manifestation of the
probability currents in the phase space\footnote{%
In much of the physics community, \textquotedblleft phase space\textquotedblright\
is a term used for the space of $x$-$p$ (coordinate and momentum).
Significantly, these variables are even/odd under time reversal. In this
paper, however, we use this term in the sense common in the dynamical
systems and climate science communities. In the cases we consider here, there is no
reason to regard the variables (e.g., temperature and volume, or two
amplitudes of a principal component analysis) as having different symmetry
under time reversal. For many in the community of statistical physics,
the familiar term in this context is \textquotedblleft configuration
space.\textquotedblright\ We will use the two terms interchangeably and
assume there is no confusion.} of the climate system.

\begin{figure}[tbp]
\centering 
\includegraphics[height=2.5in]{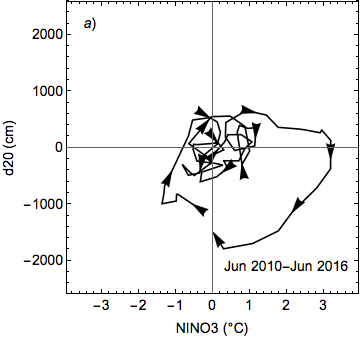} %
\includegraphics[height=2.5in]{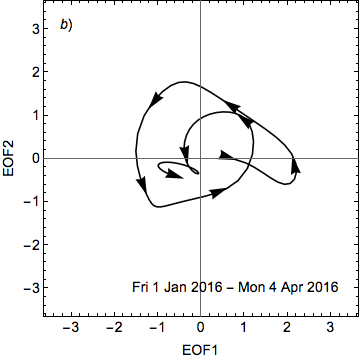}
\caption{Phase space trajectories for a) ENSO and b) the MJO}
\label{fig:ENSOMJO}
\end{figure}

Climate oscillations are often characterized by climate \textquotedblleft
indices\textquotedblright. These indices are empirically determined combinations of climate
variables, typically chosen by researchers in the subject area of interest to highlight the most important 
features of a particular phenomenon. It is not uncommon for different researchers to define different 
indices to highlight different aspects of the same complex
high-dimensional phenomenon.  Climate indices are
  commonly used to measure the amplitude of an oscillation, determine
  its power spectrum, compare oscillations among models and between
  models and observations, and for many other uses.
While it is most common to focus on a single index, sometimes
two indices are used to describe an oscillation and investigate the
trajectories of the indices in the resulting two-dimensional phase space.
Climate oscillations are then observed to have trajectories which exhibit
phase space rotation. For example, ENSO is often described in terms of the
NINO3 index, based on the spatially-averaged Sea Surface Temperature in the
eastern tropical Pacific (90$^{\circ }$W to 150$^{\circ }$W and 5$^{\circ }$%
S to 5$^{\circ }$N), and the average depth of the 20$^{\circ }$C isotherm
over the same area, which is a measure of the volume of warm water in the
tropical Pacific. The two-dimensional phase space of these indices clearly
shows the rotation characteristic of fluctuations within NESS, as seen in
Figure~\ref{fig:ENSOMJO}a (e.g. \cite{MeinenMcPhaden00,TimmermannEtAl2018}). Similar phase
space rotation is seen (Figure~\ref{fig:ENSOMJO}b) in a multivariate MJO
index \cite{WheelerHendon04} based on spatial patterns of variability of
outgoing long-wave radiation anomalies, which are a convenient observable closely related to changes in cloud cover. We propose a novel and natural
measure to quantify such rotations: the probability angular momentum (PAM).
Not only is it intimately related to the underlying probability current
loops in phase space, it is readily computed from both observations and
models, providing researchers in the climate community a new tool for model
diagnosis, validation and intercomparison.

One of the simplest classes of models that captures nonequilibrium
steady-states are Langevin models based on multivariate linear stochastic
differential equations with additive noise. Since their introduction by
Uhlenbeck and Orstein \cite{UhOr30}, these models have been 
applied to many problems in the physics community. They are also used to
model many aspects of the climate system, and,
when constructed by fitting to data, are known as linear inverse models,
e.g. \cite{LIM}. Generally assumed to have Gaussian noise
and referred to as Linear Gaussian Models (LGM), they have been successfully
used to describe a variety of climate oscillations \cite{LIM,PenlandMatrosova1998,AlexanderEtAl2008,HawkinsSuttton2009,CavanaughEtAl2015,DiasEtAl2018,stevenson2013generalized}. The simplicity of these
LGMs allows the properties of the NESS to be calculated analytically and
facilitates the presentation of PAM. We emphasize that, although we analyze
the PAM in the context of LGMs, the quantity itself is quite general, can be calculated from
observations with no assumptions about underlying models, and
captures the phase space rotation for any system regardless of the
underlying dynamics.

For the convenience of readers in the climate science community, we provide
in the next section a brief review of the role played by probability current
loops in NESSs in general and in LGMs in particular. 
Section~\ref{sec:angmom} introduces the probability angular
momentum - its average as well as its full distribution. 
In section~\ref{sec:climosc} we calculate the probability angular 
momentum of two climate oscillations: ENSO and the MJO. 
We end with a summary and outlook.

\section{Nonequilibrium Steady-states and Probability Currents}

\label{sec:ness}

The highly successful Boltzmann-Gibbs framework for
equilibrium statistical mechanics is based on a single hypothesis, that an
\textit{isolated} physical system can be found (starting with any initial 
state but waiting for a time long compared to all intrinsic relaxation 
time-scales) in any of its allowed configurations, $\{ C \}$, with equal 
probability,  i.e., the probability distribution function (pdf) being\footnote{%
Below, we will be considering time-dependent distributions, which we denote
by $P\left( C;t\right) $. The superscript ($^{\ast }$) signifies a
stationary distribution.}
$P_{iso}^{\ast }\left( C\right) \propto 1$ .
Considering two systems that can exchange energy
(or particles, or other quantities), but otherwise isolated, we arrive at a
well defined notion of thermal equilibrium and 
in particular, the Boltzmann
pdf: $P_{B}^{\ast }\propto e^{-\mathcal{H} / k_{B} T}$, where $\mathcal{H}$ is the energy of a configuration $C$. Further, the dynamical behavior
in the equilibrium state is symmetric under time-reversal: One cannot
distinguish (statistically) a movie taken of this system from another run in reverse.
However, this framework fails to describe many interesting nonequilibrium phenomena and, in particular, the climate system. One reason is clear: The climate system is driven by incoming shortwave
solar energy, balanced by outgoing long-wave radiation. 
Roughly speaking, it is coupled to two thermal reservoirs, the Sun at 
6000$^{\circ }$K and the cosmic microwave background at 3$^{\circ }$K. 
The climate system is best regarded as a (approximate) nonequilibrium
steady-state. In general, there is no simple way to find, or to hypothesize,
the probability distribution of a NESS. Though such a state is, by
definition, invariant under time \textit{translation}, it is not so under
time reversal. To make progress for this challenging problem, we may start
with a master equation\footnote{%
Though the form of our equation appears to be for continuous $t$ and
discrete $C$, it is simple to write equations for other types of variables,
e.g., continuous $t$ and $C$. Note that we have restricted ourselves to
systems evolving with time-\textit{independent} rates.} 
for the evolution of the pdf:%
\begin{equation}
\frac{\partial }{\partial t}P\left( C;t\right) = 
\sum_{C^{\prime }}\left[
W\left( C^{\prime }\rightarrow C\right) P\left( C^{\prime };t\right)
-W\left( C\rightarrow C^{\prime }\right) P\left( C;t\right) \right]  
\label{ME}
\end{equation}%
by postulating a set of (non-negative) rates, $W\left( C^{\prime
}\rightarrow C\right) $, for the system to make a transition from $C^{\prime
}$ to $C$. As probability is conserved, this equation can be regarded as a
continuity equation for the density $P$, with the terms on the right
representing net probability currents, $K\left( C^{\prime }\rightarrow
C\right) $, from $C^{\prime }$ to $C$. As the system settles into
stationarity, $P\left( C;t\right) \rightarrow P^{\ast }\left( C\right) $
(which is unique if the $W$'s allow the system to reach to all $C$'s), while 
$K$ settles into $K^{\ast }$. Of course, the sum of the $K^{\ast }$'s into
each $C$ must vanish. The principal difference between systems in thermal
equilibrium and NESS is that, in the former, \textit{every} $K^{\ast }$
vanish. A set of $W$'s which leads to such a condition, $W\left( C^{\prime
}\rightarrow C\right) P^{\ast }\left( C^{\prime }\right) =W\left(
C\rightarrow C^{\prime }\right) P^{\ast }\left( C\right) $, is said to obey
detailed balance (DB)\footnote{%
In this form, the criterion for the $W$'s to satisfy DB appears to depend on 
$P^{\ast }$. Kolmogorov \cite{Kolmo36} provided a criterion which
involves only the $W$'s.}. To model general stochastic processes, we
typically encounter $W$'s which violate DB. Those systems settle into NESS,
with some non-trivial and persistent probability currents. Being in the
stationary state, these $K^{\ast }$'s must form closed loops. Our goal is to
find observable manifestations of such persistent current loops and we will
find that these quantities are automatically \textit{odd} under time
reversal. Further details of this approach may be found in ref. \cite{ZS07}.

For a large variety of physical systems, a more restrictive version of Eqn. 
(\ref{ME}) is quite adequate, namely, the Fokker-Planck equation \cite{Riskin89}.
Specifically, suppose our configuration space consists of $N$ real variables
-- $\mathbf{x}\in \mathbb {R}^{N}$ or $x_{\alpha }$, $\alpha =1,...,N$ -- and
the only non-vanishing transitions -- $W\left( C^{\prime }\rightarrow
C\right) $ -- take the system from $C^{\prime }$ to infinitesimally nearby 
$C$'s. Then, Eqn. (\ref{ME}) reduces to%
\begin{equation}
\frac{\partial }{\partial t}P\left( \mathbf{x},t\right) =\sum_{\alpha }\frac{%
\partial }{\partial x_{\alpha }}\left[ \sum_{\beta }\frac{\partial }{%
\partial x_{\beta }}\left\{ D_{\alpha \beta }\left( \mathbf{x}\right) P\left( 
\mathbf{x},t\right) \right\} -\mu _{\alpha }\left( \mathbf{x}\right) P\left( \mathbf{x%
},t\right) \right]  \label{FP}
\end{equation}%
where $D_{\alpha \beta }$ is referred to as the diffusion tensor and $\mu
_{\alpha }$, the drift vector. The advantage of this form is the
correspondence to the continuity equation in, say, fluid dynamics: $\partial
\rho /\partial t=-\mathbf{\nabla}\cdot \mathbf{J}$, where $\rho \left( \mathbf{x}%
,t\right) $ is the fluid density field and $\mathbf{J}$, the current (density).
In this way, we identify the probability current (density) as%
\begin{equation}
K_{\alpha }=\mu _{\alpha }P-\partial _{\beta }D_{\alpha \beta }P  \label{K}
\end{equation}%
where repeated indices are summed. Similarly, in analogy with $\mathbf{J}=\rho 
\mathbf{v}$ in fluids, we identify the velocity field as%
\begin{equation}
u_{\alpha }\left( \mathbf{x},t\right) =\mu _{\alpha }\left( \mathbf{x}\right)
-\partial _{\beta }D_{\alpha \beta }\left( \mathbf{x}\right) -D_{\alpha \beta
}\left( \mathbf{x}\right) \partial _{\beta }\ln P\left( \mathbf{x},t\right).
\label{u}
\end{equation}
For the remainder of this paper, we restrict
  ourselves to the case where the diffusion matrix $D_{\alpha \beta}$ is
  independent of state of the system, $\mathbf{x}$.

Using standard techniques, this stochastic process can be recast as the more
intuitive Langevin equation%
\begin{equation}
\frac{d\mathbf{x}}{dt}=\boldsymbol{\mu}+\boldsymbol{\eta}  \label{LE}
\end{equation}%
Here, we recognize $\boldsymbol{\mu}$ as the deterministic part of this equation of
motion (and typically depends only on $\mathbf{x}$), while $\boldsymbol{\eta}$ is an
additive Gaussian noise with zero mean (i.e., $\left\langle \boldsymbol{\eta}%
\right\rangle \equiv 0$) and covariance 
$\left\langle \eta _{\alpha } ( t ) \eta_{\beta } (t ^\prime) \right\rangle 
=2D_{\alpha \beta } \delta ( t - t ^\prime) $. To be specific, we can focus on
the Ito formulation of this stochastic differential equation, e.g.,
discretizing time into steps of $\varepsilon $ and letting 
\footnote{%
Note that the discrete version of the $\delta$ in the noise correlation is a 
Kronecker delta of the time steps divided by $\varepsilon $. Note also
that there is no correlation between $\mathbf{x}\left( t\right) $ and 
$\boldsymbol{\eta}\left( t\right) $, so that 
$ < \mathbf{x}\left( t\right) \boldsymbol{\eta}\left( t\right) > \equiv 0$.}
$\mathbf{x}\left(
t+\varepsilon \right) =\mathbf{x}\left( t\right) +\varepsilon \left[ \boldsymbol{\mu}%
\left( \mathbf{x}\left( t\right) \right) +\boldsymbol{\eta}\left( t\right) \right] $ 
 
In the remainder of
this section, we will present these ideas in the context of LGMs. Being the
simplest version of (\ref{LE}) and exactly solvable \cite{Lax60,JBW03,ZS07},
they offer a clear and concise setting for us to introduce the notion 
of probability angular momentum. Further, since LGMs are frequently used in
the climate community \cite{LIM,PenlandMatrosova1998,AlexanderEtAl2008,HawkinsSuttton2009,CavanaughEtAl2015,DiasEtAl2018}, we will exploit them in describing two
examples from the global climate system, showcasing characteristics of NESS which cannot fit within the framework of thermal equilibrium.

The LGM is completely specified by two matrices,
$\mathbb A$ and $\mathbb D$, 
with \textit{constant} elements $A_{\alpha \beta }$ and 
$D_{\alpha \beta }$. The former characterizes the deterministic relaxation
into the stationary state, while stability of the system requires the real
parts of its eigenvalues to be negative. It enters Eqn. (\ref{LE}) through%
\begin{equation}
\boldsymbol{\mu}={\mathbb A\mathbf{x}}  \label{mu=Ax}
\end{equation}%
and leads to the term \textquotedblleft linear\textquotedblright\ in LGM. The
latter, also known as the diffusion matrix, describes the covariance of the
noise\footnote{%
The $D_{\alpha \beta }$ here is the same as the one introduced above, the only
difference being it is restricted to be $x$-independent in a LGM.}, and so,
must be positive definite. An alternative expression for this noise is that
its distribution is Gaussian: $\propto \exp \left\{ -
\eta _{\alpha }\Gamma ^{\alpha \beta }\eta _{\beta }/2\right\} $, where 
$\Gamma ^{\alpha \beta }$ is the matrix inverse of 
$2D_{\alpha \beta } / \varepsilon $ .
Denoting the pdf for our LGM\footnote{%
To avoid confusion, we use different notation for quantities in a LGM from
the general case, e.g., $p$ and $\mathbf{j}$ instead of $P$ and $\mathbf{K}$.} by 
$p(\mathbf{x},t)$ and the probability current density by $\mathbf{j}(\mathbf{x},t)$,
the Fokker-Planck equation becomes $\partial p/\partial t+\mathbf{\nabla}\cdot 
\mathbf{j}=0$, with 
\begin{equation}
\mathbf{j}={\mathbb A}\mathbf{x}p-{\mathbb D}\mathbf{\nabla}p
\end{equation}%
In analogy with $\mathbf{J}=\rho \mathbf{v}$ in fluid dynamics, we may identify
the velocity field as 
$\mathbf{u}={\mathbb A}\mathbf{x}-{\mathbb D}\nabla \ln p $. 
In the stationary state, the pdf, $p^{\ast }\left( \mathbf{x}\right) $, is
a Gaussian \cite{Lax60}%
\begin{equation}
p^{\ast }(\mathbf{x})=\left. e^{-\mathbf{x}^{T}{\mathbb C}_{0}^{-1}
\mathbf{x}/2}\right/ \sqrt{2 \pi \det {\mathbb C}_{0}}.
\end{equation}%
where\footnote{%
Here, $ <\mathcal{O}> ^{\ast }$ refers to the average
in the stationary state: $\int \mathcal{O}\left( \mathbf{x}\right) p^{\ast
}\left( \mathbf{x}\right) d\mathbf{x}$.} 
${\mathbb C}_{0}=\left\langle \mathbf{x}%
\mathbf{x}^{T}\right\rangle ^{\ast }$ is the covariance matrix, related to ${%
\mathbb A}$ and ${\mathbb D}$ by the generalized fluctuation
dissipation relation \cite{Lax60,JBW03} (Einstein relation):%
\begin{equation}
{\mathbb A}{\mathbb C}_{0}+{\mathbb C}_{0}{\mathbb A}%
^{T}+2{\mathbb D}=0.  \label{symPart}
\end{equation}

As presented, the LGM is able to describe systems which
settle into either thermal equilibrium or a NESS. If the matrices ${\mathbb A}
$ {and} ${\mathbb D}$ satisfy a further constraint, i.e., $
{\mathbb A}^{-1}{\mathbb D}$ being symmetric, then DB is satisfied and the
system settles into thermal equilibrium with ${\mathbb C}_{0}={%
\mathbb A}^{-1}{\mathbb D}$ and $\mathbf{j}^{\ast }\equiv 0$.
Otherwise, $\mathbf{j}^{\ast }\neq 0$ and we have a NESS. It is straightforward
to compute $\mathbf{j}^{\ast }$ which can be written as $-\left[ {\mathbb A%
}{\mathbb C}_{0}+{\mathbb D}\right] \mathbf{\nabla}p^{\ast }$.
Further, it is instructive to exploit (\ref{symPart}) and write%
\begin{equation}
\mathbf{j}^{\ast }=\frac{{\mathbb {\Lambda}}}{2}\mathbf{\nabla}p^{\ast } 
\label{j*}
\end{equation}%
where%
\begin{equation}
\mathbb{\Lambda} \equiv \mathbb C_{0}{\mathbb A}^{T}-{%
\mathbb A}{\mathbb C}_{0}  \label{Lmatrix}
\end{equation}%
Since ${\mathbb {\Lambda}}$ is manifestly antisymmetric, 
$\mathbf{\nabla} \cdot \mathbf{j}^{\ast }=0$ follows readily. Being divergence free, these
currents must form closed loops, leading to the notion of rotations and
\textquotedblleft angular momenta\textquotedblright\ - the main focus of the next section.

We end this section with a few remarks. The expression (\ref{j*}) allows us
to visualize the $\mathbf{j}^{\ast }$ field easily, as $p^{\ast }$ can be
regarded as a \textquotedblleft hill with ellipsoidal
contours\textquotedblright\ ($N-1$ dimensional sheets) to which\ $\mathbf{\nabla%
}p^{\ast }$ is perpendicular. Thus, ${\mathbb {\Lambda}}$ will generate
a vector field lying within these contours. If we focus on a two-dimensional
($N=2$) space, then the contours are ellipses and $\mathbf{j}^{\ast }$ is
tangent to them. Explicit examples of such $\mathbf{j}^{\ast }$ fields can be
found in, e.g., various figures in Ref. \cite{MMZ16,MMZ17}. Using 
$\mathbf{u}=\mathbf{K}/P $, we may associate $\mathbf{j}^{\ast }/p^{\ast }$ with a 
\textquotedblleft probability velocity field\textquotedblright\ (in the NESS) which
\textquotedblleft carries probability from one configuration ($\mathbf{x}$) to
a nearby one in preferred directions.\textquotedblright\ For the LGM, it is 
$\mathbf{u}={\mathbb  \Omega}\mathbf{x}$, where, ${\mathbb  \Omega}={\mathbb A}%
+{\mathbb D}{\mathbb C}_{0}^{-1}$ is an angular velocity,
providing us with the frequency of rotation in configuration space. Finally,
note that we can decompose ${\mathbb A}{\mathbb C}_{0}$ into
symmetric and antisymmetric parts%
\begin{equation}
-{\mathbb A}{\mathbb C}_{0}={\mathbb D}+\frac{{\mathbb \Lambda}}{2}  \label{decomp}
\end{equation}%
Below, we will discuss the significance of this decomposition.

\section{Probability Angular Momentum, its Generalization and Distribution}

\label{sec:angmom}

Probability current loops appear to be abstract concepts;
how are they manifested in physical observables? Given that we expect
rotations, angles and angular velocities (in configuration space) come
naturally to mind \cite{RussellBlythe13}. However, there are disadvantages to these quantities, such as
dependence on the choice of origin for $ \mathbf{x}$ and
singular properties when the trajectory $\mathbf{x}\left( t\right) $ come close
to this origin. We will argue that the analog of angular momentum is a
better choice, related to many familiar quantities which are normally used
to characterize a time series of observables.

In classical mechanics, the angular momentum associated with a point
particle of mass $m$ at $\mathbf{r}\left( t\right) $ moving with velocity $\mathbf{%
v}\left( t\right) $ is $\mathbf{L} \left( t\right) =\mathbf{r}\times m\mathbf{v}$. Clearly, for a
collection of such particles, mass $m_{i}$ at $\mathbf{r}_{i}$ moving with
velocity $\mathbf{v}_{i}$, the total angular momentum is just $\sum_{i}\mathbf{r}%
_{i}\times \mathbf{v}_{i}m_{i}$ . \ The next step is to generalize to a
continuous distribution described by a mass density $\rho (\mathbf{r},t)$ and a
velocity field $\mathbf{v}(\mathbf{r},t)$, such as a fluid from an Eulerian
perspective. From these, we construct $\,\mathbf{r}\times \mathbf{v}\rho d\mathbf{r}$
and regard $\mathbf{L}(\mathbf{r},t)=\mathbf{r}\times \mathbf{v}(\mathbf{r},t)\rho (\mathbf{r}%
,t)$ as the angular momentum \textit{density}. The \textit{total} angular
momentum of the entire distribution is\footnote{%
Note that we use the same letter for both the density of a quantity and the
total, the former having an additional argument, $\mathbf{x}$.} $\mathbf{L}(t)=
\int d\mathbf{r}\,\rho (\mathbf{r},t)\,\mathbf{r}\times \mathbf{v}(\mathbf{r},t)$ , 
or substituting $\mathbf{J}$ for $\rho \mathbf{v}$,%
\begin{equation}
\mathbf{L}(t)=\int d\mathbf{r}~\mathbf{r}\times \mathbf{J}(\mathbf{r},t)  \label{L in CM}
\end{equation}%
From here, we propose the \textit{probability angular momentum} in
configuration space as a straightforward analogue, by letting $\mathbf{r}%
\rightarrow \mathbf{x}$, $\mathbf{v}\rightarrow \mathbf{u}$, $\rho \rightarrow P$,
and $\mathbf{J}\rightarrow \mathbf{K}$. Now, in $N$ dimensions, the PAM is no
longer a (pseudo\mbox{-})vector but a (pseudo\mbox{-})tensor, ${\mathbb  L}$, with components%
\begin{equation}
L_{\alpha \beta }\left( t\right) =\int d\mathbf{x}~\left( x_{\alpha }K_{\beta }(%
\mathbf{x},t)-x_{\beta }K_{\alpha }(\mathbf{x},t)\right)  \label{PAM}
\end{equation}%
The integrand here can be regarded as the \textit{probability angular
momentum density}%
\begin{equation}
L_{\alpha \beta }(\mathbf{x},t)=x_{\alpha }K_{\beta }(\mathbf{x},t)-x_{\beta
}K_{\alpha }(\mathbf{x},t)  \label{L density}
\end{equation}%
Since $P$ is normalized to unity, the analogue of total mass is simply
unity. Thus, if the units of all components of $\mathbf{x}$ are the same, $%
\left[ x\right] $, then the units of the PAM (\ref{PAM}) are $\left[ x\right]
^{2}\left[ t\right] ^{-1}$ -- the same as diffusion (cf. below).
In a NESS, we have ${\mathbb  L}^{\ast }$, i.e., $L_{\alpha \beta }^{\ast
}=\int d\mathbf{x}~\left( x_{\alpha }K_{\beta }^{\ast }-x_{\beta }K_{\alpha
}^{\ast }\right) $ and note one of the salient features of using angular
momenta instead of angles: Since $\mathbf{\nabla}\cdot \mathbf{K}^{\ast }=0$,
${\mathbb  L}^{\ast }$ is \textit{independent }of the choice of the
origin for $\mathbf{x} $.

Turning to the LGM (\ref{LE},\ref{mu=Ax}) in the steady state, the elements of ${\mathbb  L}^{\ast }$ are
\begin{equation*}
L_{\alpha \beta }^{\ast }=\int d\mathbf{x}\left( x_{\alpha }j_{\beta }^{\ast
}-x_{\beta }j_{\alpha }^{\ast }\right)
\end{equation*}%
Substituting (\ref{j*}) and integrating by parts, we find a very simple result for the LGM:
\begin{equation}
{\mathbb L}^{\ast }={\mathbb {\Lambda}} \label{LGM-L}
\end{equation}%
Of course, since $\mathbf{j}\propto p$, with components of $\mathbf{x}$ as
coefficients, ${\mathbb  L}^{\ast }$ is just linear combinations of the
two-point correlation ${\mathbb C}_{0}$. Indeed, it is just twice the
antisymmetric part of $-{\mathbb A}{\mathbb C}_{0}$, while Eqn. (%
\ref{decomp}) puts the PAM on the \emph{same footing }as diffusion\emph{.}
It is not an accident that, as noted above, the units of the PAM are those
of diffusion. While we are yet to understand the deeper significance underlying
angular momenta and diffusion being part of one quantity 
(associated with the space of probability density), 
we can regard this relationship as another
salient feature of using the PAM instead of angles to characterize rotation
in statistical mechanics.

Before continuing onto generalizations and distributions of the PAM, we
present a brief summary of an explicitly analyzied, simple LGM \cite%
{ZS07,UGa16} which may provide a helpful setting for more complex and
realistic phenomena. Consider a system with just two degrees of freedom ($%
x_{1,2}$), specifically, two coupled simple harmonic oscillators governed by a
Hamiltonian, $\mathcal{H} = \left[ k_{1}x_{1}^{2}+k_{2}x_{2}^{2}+k_{\times
}\left( x_{1}-x_{2}\right) ^{2}\right] /2$, each immersed in its own thermal
bath of temperature $T_{1,2}$. This system has been previously
  studied by Ciliberto, et al.  \cite{CilibertoEtAl2013}
In the low-mass over-damped limit, this
system is naturally described by Langevin equations (Boltzmann's constant
absorbed into $T$): $dx_{\alpha }/dt=-\lambda _{\alpha }\left( \partial 
\mathcal{H}/\partial \xi _{\alpha }\right) +\eta _{\alpha }$, with $%
\left\langle \eta \right\rangle =0$ and $\left\langle \eta _{\alpha }(t)\eta
_{\beta }(t^{\prime })\right\rangle =2\lambda _{\alpha }T_{\alpha }\delta
_{\alpha \beta }\delta (t-t^{\prime })$. Note that, in the absence of
coupling ($k_{\times }=0$) or $T_{1}-T_{2}=0$, this system will settle into
thermal equilibrium. Otherwise, it is precisely an LGM with (\ref{mu=Ax}) in
(\ref{LE}). The $2\times 2$ matrices ${\mathbb A}$ {and} ${\mathbb D}$ 
are readily identified, while ${\mathbb C}_{0}$ and 
${\mathbb {\Lambda}}$ can be easily computed. In the latter, there is a single
independent component, say, $\Lambda_{12}^{\ast }$, which is proportional to 
$k_{\times }\left( T_{1}-T_{2}\right) $. Thus, we see that 
$L_{12}^{\ast } = \Lambda_{12}^{\ast }$
is non-zero if and only if $k_{\times }\neq 0$ and $T_{1}\neq T_{2}$, and
the system then settles into a NESS.

So far, our study of the PAM has led us to two point correlations at
(almost) equal times, since $\mathbf{u}=d\mathbf{x}/dt=\underset{\varepsilon
\rightarrow 0}{\lim }\left[ \mathbf{x}\left( t+\varepsilon \right) -\mathbf{x}%
\left( t\right) \right] /\varepsilon $. There is a natural generalization to
correlations at arbitrary times $t\neq t^{\prime }$. Though rarely
considered in classical mechanics of point particles, this generalization
takes the form%
\begin{equation}
\mathbf{A}\left( t,t^{\prime }\right) \equiv m\mathbf{r}\left( t\right) \times 
\mathbf{r}\left( t^{\prime }\right)  \label{A}
\end{equation}%
for any given trajectory $\mathbf{r}\left( t\right) $. Note that the magnitude $%
\left\vert \mathbf{A}\right\vert $ is the area of a parallelogram spanned by
the two $\mathbf{r}$'s (related to the area in Kepler's second law).
As $t^{\prime }\rightarrow t_{+}$, $\mathbf{L}$ is recovered: $\mathbf{L}%
\left( t\right) =\left. \partial \mathbf{A}\left( t,t^{\prime }\right)
/\partial t^{\prime }\right\vert _{t^{\prime }=t}$. The statistical 
mechanics analog of $\mathbf{A}$ is Levy's stochastic area\footnote{%
Introduced in 1940 by Levy \cite{Levy40}, this concept was subsequently
developed in a series of articles. For a modern review and further 
generalizations, see, e.g., Ref. \cite{HelmesSchwane83}. Most recently, 
it has been exploited in the context of noisy couple RC 
circuits \cite{GNT17,GNT19}.}. 
Here, we consider the two point correlation function at
unequal times, $C_{\alpha \beta }\left( t,t^{\prime }\right) $. To be
precise, the definition%
\begin{equation}
C_{\alpha \beta }\left( t,t^{\prime }\right) \equiv \left\langle x_{\alpha
}(t)x_{\beta }(t^{\prime })\right\rangle =\int x_{\alpha }x_{\beta }^{\prime
}P\left( \mathbf{x},t;\mathbf{x}^{\prime },t^{\prime }\right) d\mathbf{x}\,d\mathbf{x}%
^{\prime }  \label{2pf}
\end{equation}%
requires the \textit{joint} probability distribution:
$P\left( \mathbf{x},t;\mathbf{x}^{\prime },t^{\prime }\right) =P\left( \mathbf{x},t\right) G\left( \mathbf{x}^{\prime }-\mathbf{x},t^{\prime }-t\right) $. 
Here, $G\left( \mathbf{\xi},\tau \right) $ is the time dependent solution to the
Fokker-Planck equation (\ref{FP}), subjected to the initial condition 
$G\left( \mathbf{\xi},0\right) =\delta \left( \mathbf{\xi}\right) $. The analog of 
$\mathbf{A}\left( t,t^{\prime }\right) $, i.e., the generalization of 
${L_{\alpha \beta }(t)}$, is just the antisymmetric combination 
\begin{equation}
\tilde{C}_{\alpha \beta }\left( t,t^{\prime }\right) \equiv C_{\alpha \beta
}\left( t,t^{\prime }\right) -C_{\beta \alpha }\left( t,t^{\prime }\right)
\label{C-tilde}
\end{equation}%
so that ${L_{\alpha \beta }}\left( t\right) =\left. d\tilde{C}_{\alpha \beta
}\left( t,t^{\prime }\right) /dt^{\prime }\right\vert _{t^{\prime }=t}$. By
construction, $\tilde{C}$ is odd under $t\Leftrightarrow t^{\prime }$: $%
\tilde{C}_{\alpha \beta }\left( t,t^{\prime }\right) =-\tilde{C}_{\alpha
\beta }\left( t^{\prime },t\right) $, a property related to the violation of
time reversal symmetry. Of course, in the steady state $\tilde{C}{^{\ast }}$
is stationary, so it depends only on the difference $\tau =t^{\prime }-t$.
This formulation parallels that of relative dispersion in
fluids \cite{Batchelor52,DukowiczSmith97,LaCasce08}.

Clearly, much more information about the dynamics of our system is encoded
in $\tilde{C}^{\ast }$ than in $L^{\ast }$ and, naturally, computing these
quantities theoretically is difficult in general. However, as the LGM is completely
specified by the matrices ${\mathbb A}$ and ${\mathbb D}$, it can
be found analytically. In particular, the time-lagged covariance matrix ${%
\mathbb C}_{\tau }$ (i.e., $C_{\alpha \beta }^{\ast }\left( t,t+\tau
\right) $) is related simply to ${\mathbb C}_{0}$: 
\begin{equation}
{\mathbb C}_{\tau }=\left\langle \mathbf{x}(t+\tau )\mathbf{x}%
^{T}(t)\right\rangle ^{\ast }=e^{{\mathbb A}\tau }{\mathbb C}_{0}
\label{eq:Clag}
\end{equation}%
As expected, it is independent of $t$, due to time translational invariance
of a steady-state. Instead of ${\mathbb A}$ and ${\mathbb D}$, the
LGM can be specified alternatively by these steady-state covariance
matrices ${\mathbb C}_{0}, {\mathbb C}_{\tau }$. The advantage is that this representation is useful for
constructing empirical models from data as discussed below. Further, note
that $\left. d{\mathbb C}_{\tau }/d\tau \right\vert _{\tau =0}={%
\mathbb A\mathbb C}_{0}$ contains the full information contained
in ${\mathbb D}$ and ${\mathbb \Lambda }$. Thus, an alternative
perspective of an LGM is to specify it by ${\mathbb C}_{0}$, ${%
\mathbb D}$ and ${\mathbb \Lambda}$. The advantage of this
representation is that all possible systems can be grouped into families
with the same ${\mathbb C}_{0}$ and ${\mathbb D}$ but different 
${\mathbb \Lambda}$. Since ${\mathbb D}$ is positive definite, one can always transform to coordinates where ${\mathbb D}$ is diagonal \cite{JBW03}. In these coordinates, diffusion is solely in the radial direction in phase space. Then, phase space rotation is completely captured by ${\mathbb \Lambda}$. To emphasize, only one member of the family (the one with 
${\mathbb \Lambda}=0$) corresponds to a system in thermal equilibrium.
All other members represent nonequilibrium systems (with the same pdf and
diffusion) with non-trivial PAM. We will not pursue the study of ${%
\mathbb C}_{\tau }$ further here, but examples of its behavior can be
found in other contexts \cite{SZ14,MMZ16,MMZ17}.

In the final paragraphs of this section, we consider another important
aspect of the PAM, namely, its full distribution. Focusing only on a NESS,
we observe the system for a length of time and record a single trajectory: $%
\mathbf{x}_{obs}\left( t\right) $. Of course, physical data
such as in climate science would be observed at typically integer multiples of some finite
time step, $\varepsilon $. Then, instead of the continuous velocity, 
we would use the finite difference approximation:
$d\mathbf{x}/dt\approx \left[ \mathbf{x}(t+\varepsilon )-\mathbf{x} (t)\right] /\varepsilon $.
From such time series, we can construct another series for 
\begin{equation}
{\mathbb  L}_{obs}\left( t\right) =\mathbf{x}_{obs}\left( t\right) \wedge 
\mathbf{x}_{obs}\left( t+\varepsilon \right) /\varepsilon   \label{PAM(t)}
\end{equation}%
where we have used the wedge product to denote the antisymmetric part of the
tensor product. Note that the extra term in the discrete approximation for the velocity plays no role, as 
$\mathbf{x}_{obs}\left( t\right) \wedge \mathbf{x}_{obs}\left( t\right)
\equiv 0$. Meanwhile, under the assumption of ergodicity, various
statistical quantities can be computed from a \textit{time average} over the
trajectory (denoted by an overline) instead of the ensemble averages
discussed above, e.g., 
\begin{equation*}
\overline{{\mathbb  L}_{obs}}={\mathbb  L}^{\ast }
\end{equation*}%
In addition to the average over the time series, we can construct a
histogram, which approximates the full distribution of the quantity
involved. For convenience, let us focus on a two-dimensional configuration
space (which can be a subspace of a higher dimensional space) and, dropping
the subscript $_{obs}$, we will simply label the observed values as 
$x_{1}\left( t\right), x_{2}\left( t\right)$. Then, the PAM is characterized
by a single (independent) quantity, say, the $1$-$2$ element of ${\mathbb  L%
}_{obs}$. We denote that time series by%
\begin{equation*}
\mathcal{L}\left( t\right) \equiv \left( x_{1}(t)x_{2}(t+\varepsilon
)-x_{2}(t)x_{1}(t+\varepsilon )\right) /\varepsilon 
\end{equation*}%
and the associated histogram by $H\left( \mathcal{L}\right) $. Note that, in
general, $\mathcal{L}\left( t\right) $ will appear with both signs and the
support of $H$ is over the entire line $\left( -\infty ,\infty \right) $.
Normalizing $H$ provides us with a pdf, which can be compared to the
theoretical expression 
\begin{equation*}
f\left( \mathcal{L}\right) =\int \delta \left( \mathcal{L}-\frac{%
x_{1}x_{2}^{\prime }-x_{2}x_{1}^{\prime }}{\varepsilon }\right) P^{\ast
}\left( \mathbf{x},0;\mathbf{x}^{\prime },\varepsilon \right) d\mathbf{x}\,d\mathbf{x}%
^{\prime }
\end{equation*}%
where $P^{\ast }\left( \mathbf{x},0;\mathbf{x}^{\prime },\varepsilon \right)
=p^{\ast }\left( \mathbf{x}\right) G\left( \mathbf{x}^{\prime }-\mathbf{x}%
,\varepsilon \right) $ is the joint probability in NESS. In the LGM, $G$ is
also Gaussian, like $p^{\ast }$, so that the Fourier transform of $f$ 
\begin{equation}
\hat{f}(\phi )=\int f(\mathcal{L})e^{-i\mathcal{L}\phi }d\mathcal{L}
\end{equation}%
involves only Gaussian integrals and can be computed exactly. The technical
details are quite involved and will be deferred to another publication. Let
us summarize the main results here.

\begin{itemize}
\item $1/\hat{f}(\phi )$ is the square root of the determinant of the
matrix appearing in the Gaussian.

\item As the matrix is $4\times 4$, $1/\hat{f}^2$ is a quartic polynomial in $\phi $.

\item The singularities of $\hat{f}(\phi )$ are branch points. Located at the roots of the quartic, they lie on both sides of the real axis, with those nearest to the real axis controlling the large 
$\mathcal{L}$ asymptotic (exponential) decay of $f(\mathcal{L})$.

\item The parameters of the quartic come from the defining matrices of
the LGM: either the pair (${\mathbb A}$,${\mathbb D}$), or the set
(${\mathbb C}_{0}$,${\mathbb D}$,${\mathbb \Lambda}$). For
our $N=2$ case, it is more convenient to use the latter set, as the first
two matrices are real symmetric. Meanwhile, let us write ${\mathbb \Lambda}$ as $\left( 
\begin{array}{cc}
0 & \ell  \\ 
-\ell  & 0%
\end{array}%
\right) $ so that \textit{all} the DB violating aspects of this LGM are
contained in a single parameter: $\ell $. 

\item As expected, in the $\ell =0$ case, $1/\hat{f}^2$ is quadratic in 
$\phi ^{2}$. Thus, the distribution $f$ is symmetric\ in $\mathcal{L}$
and leads to $\left\langle \mathcal{L}\right\rangle \equiv 0$. 

\item  $\hat{f}$ is 
of the form $1-i\phi \ell +O\left( \phi ^{2}\right) $ and so, 
$\left\langle \mathcal{L}\right\rangle =\left. id\hat{f}/d\phi \right\vert _{\phi =0}$
is just $\ell $, confirming Eqn. (\ref{LGM-L}).

\item Note that $f$ is not $\delta $ distributed and has a finite variance. 
For that, we need 
$\left\langle \mathcal{L}^{2}\right\rangle =\left. -d^{2}\hat{f}/d\phi^{2}
\right\vert _{\phi =0}$. In particular, even with $\ell=0$, this variance,
which we denote by $\sigma _{0}^{2}$, is non-trivial. The physics
is clear: A trajectory for a system in an equilibrium steady state is just
as likely to rotate one way as the other, respecting time reversal symmetry.
The typical values of these rotations are $O\left( \sigma _{0}\right) $,
associated with both the damping and the noise (${\mathbb A}$ and 
${\mathbb D}$).

\item For $ \ell \ne 0$, we find 
$\left\langle \mathcal{L}^{2}\right\rangle =\sigma _{0}^{2}+2\ell ^{2}$. 
As a result, we arrive at a simple expression%
\begin{equation*}
\sigma _{\ell }^{2}\equiv \left\langle \mathcal{L}^{2}\right\rangle
-\left\langle \mathcal{L}\right\rangle ^{2}=\sigma _{0}^{2}+\ell ^{2}
\end{equation*}%
for the variance of the distribution $f$ for systems in NESS. This leads to
an important ratio%
\begin{equation}
\frac{\left\langle \mathcal{L}\right\rangle }{\sigma _{\ell }}=\sqrt{\frac{%
\ell ^{2}}{\sigma _{0}^{2}+\ell ^{2}}}  \label{eq:error}
\end{equation}%
which implies the following caution. If a trajectory with finite time steps
is used to find averages and standard deviations of probability angular
momenta, and if a NESS system is well described by an LGM, then 
$\left\langle \mathcal{L}\right\rangle $ can never exceed $\sigma _{\ell }$.
Thus, we must examine the statistics of the full pdf in order to come to a
meaningful conclusion on whether a nonzero time average $\mathcal{\bar{L}}$
is significant or not. In contrast, stochastic processes with prominent
rotational aspects (e.g., noisy limit cycles) are not subjected to the
limitations shown here \cite{UGa16}. A comprehensive discussion is beyond
the scope of this work and will be presented elsewhere.
\end{itemize}

Within the context of the LGM, we presented a complete analytic description
of various aspects of the PAM. For systems that display prominent rotations,
there is little need to identify them as NESS. But, there are many cases
where the trajectories in configuration space display subtle rotations,
hidden behind a substantial amount of noise. For data from computer
simulations or observations, we are necessarily restricted to a finite times
series of discrete points. From such a series, we may construct ${%
\mathbb  L}_{obs}\left( t\right) $ according to (\ref{PAM(t)}) and
compute the time average $\overline{{\mathbb  L}_{obs}}$. A non-zero
value is clear signal of time reversal violation, so that the system cannot
be regarded as \textquotedblleft in thermal equilibrium.\textquotedblright\
We should be cautious, however, since the variance of ${\mathbb  L}%
_{obs}\left( t\right) $ is typically non-trivial, even in equilibrium. For
an LGM with just two variables, we are able to find exact analytic results,
so that these remarks rest on a sound quantitative foundation. In
particular, we computed the full distribution of $\mathcal{L}$ (the single
independent quantity associated with ${\mathbb  L}$) and showed that it is
non-vanishing on both sides of $\mathcal{L}=0$, while a non-zero average
depends on the subtle asymmetry of this distribution. Given a times series, $%
\mathbf{x}_{obs}\left( t\right) $, a histogram can be compiled for the
associated $\mathcal{L}\left( t\right) $, and we can compare that to the
theoretical distribution. Such a comparison provides a further criterion,
beyond fitting a Gaussian to the histogram of $\mathbf{x}_{obs}$, for whether
the data can be adequately described by an LGM.

Finally, we address a natural question, namely, how do we interpret the sign
of the PAM. Our proposal is that, especially in cases of \textquotedblleft
subtle\ displays\textquotedblright\ of this rotation, one of the variables
is the \textquotedblleft driver\textquotedblright\ with the other being the
\textquotedblleft follower,\textquotedblright\ much like the increase of
prey populations \textquotedblleft drives\textquotedblright\ the increase in
the numbers of predators. In physical systems, the sign of $\mathcal{\bar{L}}
$ may point us to more tractable underlying causes of this \textquotedblleft
driver-follower\textquotedblright\ behavior, a key characteristic of NESS.  In the paleoclimate literature, a similar interpretation is used with ``phase wheels'' used to indicate the ordering of phenomena and likely causality under cyclical forcing \cite{imbrie1993structure}, here the cycles result from the PAM.  
In the next section, we apply these finding to two prominent examples of natural variability
in the climate system. Of course, like most physical systems, the dynamics here
are far from being linear and the stochastics are more complex than additive
Gaussian noise. Nonetheless, for reasons that are not fully understood, it is often the case that some aspects of climate oscillations about a steady-state are skillfully modeled by LGMs.

\section{Example Climate Oscillations: ENSO and MJO}

\label{sec:climosc}

Climate oscillations are preferred spatio-temporal patterns of natural
variability of the climate system. 
These climate oscillations are preferred in the
  sense that they represent some of the most significant variability about the
time-average climate state, and are further important as they have
significant human impacts.
Each oscillation has a typical fairly  
narrow range of timescales
and has a large projection onto different, relatively small subspaces of the
massively high-dimensional phase space of the entire climate system. Climate
oscillations are quantified with climate indices: functions of subsets of
climate variables, filtered to the specific spatio-temporal scales of the
pattern, and empirically developed to capture the dominant features of a
climate oscillation with one, two, or a few scalar quantities. Generally, 
a variety of sets of climate indices can be used to quantify a climate 
oscillation with mostly consistent results \cite{wolter2011nino}.  Oscillations
often have many coevolving indices, each of which highlights a different
aspect of the complex pattern.

The El-Ni\~no Southern Oscillation (ENSO) has its largest projection in the
tropical Pacific region, dominating ocean temperatures, the location of
atmospheric convection and precipitation, and the atmospheric Walker circulation. Individual ENSO events 
(as defined by indices retaining a value beyond a critical threshold) 
persist for roughly 9 months and the time between repeated events is on
the order of 2 to 7 years. We adopt the common description of ENSO in terms of the NINO3 index
(NINO3) and the depth of the 20$^\circ$C isotherm in the tropical Pacific
(d20), which roughly indicates the depth of the thermocline which is the 
region below the surface layer where temperatures begin rapidly decreasing 
toward abyssal values. ENSO index data is publicly available from a number of sources. Here
we use the data from the KNMI Climate Explorer \cite{TrouetVan-Oldenborgh13}. 
The data used here are monthly averages of observations and extends from
1960 to 2016.

The Madden-Julien Oscillation (MJO) is an eastward-moving pattern that has
its largest projection on tropical rainfall, convection, and outgoing
long-wave radiation. Its timescale is weeks to months. We will describe the
MJO in terms of the so-called Original MJO Index (OMI) which is a
two-dimensional index representing the principal components of the first two
empirical orthogonal functions (EOFs) of filtered outgoing long-wave
radiation between $20^{\circ}$N and $20^{\circ}$S \cite{kiladis2014comparison}. Like ENSO data, MJO data is 
publicly available from a number of
sources. Here we use the daily data from NOAA's Earth System Radiation Lab 1/1/1979 - 4/26/2016.

The units of phase space for climate indices can be unintuitive. The units
of probability angular momentum are the same as the units of diffusion,
length$^2$/time. The two ENSO indices, however, have different units of
``length''. NINO3 is a temperature anomaly and has units of $^\circ$C, while
d20 is a depth anomaly and has units of cm. As a result, in the NINO3-d20
phase space with monthly data the probability angular momentum has units of 
$^\circ$C cm/month\footnote{%
Note that ``diffusion'' in this case also carries these units, as it is 
the noise covariance matrix.}. Sometimes one uses indices that have been 
scaled by their standard deviation resulting in indices which are unitless 
(Mahalanobis distance). Then, the probability angular momentum would have 
units of 1/time.

Climate oscillations can be modeled by LGMs of the form 
(\ref{LE} with \ref{mu=Ax}) through a process called linear inverse modeling. 
A multivariate time series is used to construct the steady-state covariance 
${\mathbb C}_0$ and the time-lagged covariance ${\mathbb C}_\tau$.
The time lag is empirically chosen to capture the timescale of the climate
oscillation of interest. Then Eq.~\eqref{eq:Clag} is used to compute ${%
\mathbb A}$ and Eq.~\eqref{decomp} or \eqref{symPart} determines ${\mathbb D}$. There
is no guarantee that this procedure will result in a stable SDE, but it
often works surprisingly well, e.g., 
\cite{WinklerNewman01,KirtmanShukla02,HawkinsSutton09,AlexanderEtAl2008}. 
Here
we use linear inverse modeling to construct two-dimensional LGMs for 
ENSO and the MJO from the time series of their indices.

\begin{figure}[tbp]
\centering 
\includegraphics[height=2.5in]{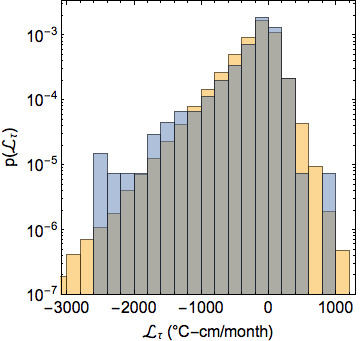} %
\includegraphics[height=2.5in]{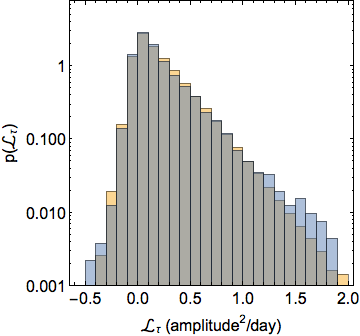}
\caption{Pdfs of the finite-time probability angular momentum for
observations and a linear Gaussian model fit to the observations. a) ENSO using
monthly data, and b) MJO using daily data. Gray indicates regions where the
two pdfs overlap, yellow indicates regions where the model pdf is larger
than the observation pdf, and blue indicates regions where the observation
pdf is larger than the model pdf.}
\label{fig:ENSOMJO2}
\end{figure}

The pdf $f\left( \mathcal{L}_{\tau }\right) $ can be calculated directly
from the observed time series as well as theoretically from the LGMs. 
The pdfs are strongly asymmetric and have exponential tails. The
asymmetry leads to the total probability angular momentum of the
steady-state, $\left\langle \mathcal{L_{\tau }}\right\rangle $, being
nonzero despite the most likely value (mode) being zero. The two methods 
of generating a pdf (measurement of $f\left( \mathcal{L}_{\tau }\right) $ 
versus LGM based on ${\mathbb A}, {\mathbb D}$) agree surprisingly well 
for the ENSO and MJO cases, despite the underlying complexity of these 
phenomena and their quantifying indices. The linear inverse 
modeling procedure produces, by construction, a LGM that has same average
$\left\langle \mathcal{L_{\tau }}\right\rangle $ as the data. Thus, we
rely on the excellent agreement between the full pdf's (Fig.~\ref{fig:ENSOMJO2})
to conclude that measurements of $f\left( \mathcal{L}_{\tau }\right) $ are robust 
and that they can be used to verify models (both LGMs and more complex climate
models) as well as for model intercomparison. Climate
  phenomena represent complex extremely high-dimensional dynamical
  systems. The probability angular momentum provides a new quantity
  based on the persistent probability currents in nonequilibrium
  steady-states to quantify the fluctuations about the time mean state.
The two examples shown here were
minimal, involving only two degrees of freedom, but the approach and
equations presented here can be applied to more detailed systems as well,
keeping \eqref{eq:error} as a guide for how many probability angular momenta
(involving various pairs of axes) are reliable. 

\section{Summary and Outlook}

In this work, we study a principal characteristic of nonequilibrium
stationary states, namely, \textit{persistent} probability current loops. Of
the many observable consequences, we propose to focus on a particularly
convenient quantity: the probability angular momentum (in analogy with
angular momenta associated with circulating fluids). Directly related to
correlation functions at unequal times, it can be used to characterize any
statistical system in NESS. Here, we provide an illustration in the context
of climate science, i.e., the two \textquotedblleft
oscillatory\textquotedblright\ phenomena, ENSO and MJO.

Exploiting the parallel with fluid dynamics, we regard the probability
density $P\left( \mathbf{x},t\right) $ (in the space of configurations of a
statistical system, or phase space) as a fluid density $\rho \left( \mathbf{r}%
,t\right) $ (in ordinary 3-dimensional space). Similarly, we draw the
parallel between current \textit{densities} $\mathbf{K}\left( \mathbf{x},t\right) $
and $\mathbf{J}\left( \mathbf{r},t\right) $, as well as velocity fields $\mathbf{u}=%
\mathbf{K}/P$ and $\mathbf{v}=\mathbf{J}/\rho $. From these, we propose to study the 
\textit{probability angular momentum density}: $\mathbb{L}\left( \mathbf{x}%
,t\right) \equiv \mathbf{x}\wedge \mathbf{u}\,P$. After our system settles into a
stationary state, described by distribution $P^{\ast }\left( \mathbf{x}\right) $%
, there will be non-trivial steady $\mathbf{K}^{\ast }$, provided it is a NESS.
These current loops lead naturally to the concept of the (total) probability
angular momentum: $\mathbb{L}^{\ast }=\int d\mathbf{x}~\mathbb{L}^{\ast }\left( \mathbf{x}%
\right) =\left\langle \mathbf{x}\wedge \mathbf{u}\right\rangle ^{\ast }$. To
emphasize, this quantity must vanish for a system in thermal equilibrium, in
which $\mathbf{K}^{\ast }\equiv 0$ and thus, serves as a quantitative measure
for the nonequilibrium characteristics of a stationary system. Since $\mathbf{u}
$ is associated with $\mathbf{x}$ at an infinitessimally later time, $\mathbb{L}%
^{\ast }$ can be generalized to two point correlation functions at arbitrary
unequal times $\left\langle \mathbf{x}(t)\wedge \mathbf{x}(t+\tau )\right\rangle
^{\ast }/\tau $. Noting that $\mathbb{L}^{\ast }$ and $\mathbb{D}$ share the
same units, we find an intimate relationship between angular momentum and
diffusion, the deeper significance of which is yet to be explored. In
general, we argue that the probability angular momentum plays a central role
in any study of the nonequilibrium fluctuations in a NESS.

If a system evolving in NESS is observed, a single trajectory would be
recorded at typically integer multiples of some time step $\varepsilon $: 
$\mathbf{x}_{obs}\left( n\varepsilon \right) $. From this, we can construct a
time series for $\mathbb{L}_{obs}\left( n\varepsilon \right) =\mathbf{x}%
(n\varepsilon )\wedge \mathbf{x}(n\varepsilon +\varepsilon )/\varepsilon $.
Invoking ergodicity, we expect the time average of $\mathbb{L}_{obs}$ to be 
$\mathbb{L}^{\ast }$.\ Further, we can study its variations and compile a
histogram and study the full distribution of $\mathbb{L}_{obs}$. In general,
the variance is non-trivial, even if the time average vanishes. As a result,
some care is needed when analyzing model simulations or physical data.
Finally, we provide an analytically tractable case for exploring these
ideas, namely, the linear Gaussian model. In particular, explicit results
are found for the general case of an LGM with just two variables (in which 
$\mathbb{L}$ is specified by a single quantity $\mathcal{L}$).

To illustrate how this approach may be applied to climate studies, we
consider certain aspects of the ENSO and MJO phenomena. Using just two
indices from the physical data in each case, we show that the histograms of 
$\mathcal{L}$ are quite asymmetric, and so, there is no doubt that the
averages $\left\langle \mathcal{L}\right\rangle $ are non-trivial. Further,
we construct LGMs using the time lagged covariance from data, and found that
its predictions are in excellent agreement with the histograms. Such
analysis gives us confidence that the LGMs capture the essentials of this
aspect of the NESS. 
In the climate context, it is important to note that the predictions 
and projections for future climate states rely on the persistent 
probability currents (Fig. 1). When LGMs are used for forecasting 
\cite{AlexanderEtAl2008}, 
it is critical to match the probability 
currents as observed which are, as emphasized here, a source of long 
timescale predictability. 
Our conclusion is that probability angular momenta
provide a valuable and novel route to study the time reversal violating
aspects of not only our climate system but also \textit{all} systems driven
far from equilibrium in general.

Naturally, many new questions arise. One line of questions follows the
applications to the climate system. Needless to say, there are many, many
more climate phenomena to which we can apply this type of analysis in
addition to the illustrations here -- the North Atlantic Oscillation,
Southern and Northern Annular Modes, variability of the Oceanic Meridional
Overturning Circulation, etc. The accuracy of models of these phenomena can
usefully be constrained by evaluating the probability angular momenta of the
model versus those of observations. The other line of pursuit is in the
realm of theory. Many issues related to the PAM within the context of the
LGM remain to be explored further. A prime example is the behavior of the
time-lagged correlation $\left\langle \mathbf{x}(0)\wedge \mathbf{x}(\tau
)\right\rangle ^{\ast }$ as a function of $\tau $, as well as their
associated distributions. General considerations and a few preliminary investigations for specific
systems \cite{SZ14,MMZ16} show that it rises to a maximum before decaying
exponentially. What is the physics behind this peak? A comprehensive study,
valid for all LGMs in arbitrary dimensional phase space, would be valuable.
What are the key characteristics of the LGM that can lead us to predict
which are the \textquotedblleft driving variables\textquotedblright\ and
which are the \textquotedblleft followers?\textquotedblright\ Beyond these
questions associated with the probability currents, loops, angular momenta,
and rotations in phase space, we should consider their implications in a
wider context. Can such considerations lead us to a quantity, or quantities,
beyond the twin pillars of equilibrium statistical mechanics: energy and
entropy? From these steps, we may find hints towards formulating a framework
for preferred fluctuations in non-equilibrium statistical mechanics and the
climate system.

\begin{acknowledgements}
This work was supported in part by NSF INSPIRE Award \#1245944 and NSF
DMR \#1507371. Work of JBW was partially carried out during a stay at
the Institute for Marine and Atmospheric research Utrecht (Utrecht
University, NL) which was supported by the Netherlands Centre for Earth
System Science. Three of us (JBW, BF-K, RKPZ) are grateful for the
hospitality of the MPIPKS, where some of this work was carried out during a
workshop \textit{Climate Fluctuations and Non-Equilibrium Statistical
Mechanics: an Interdisciplinary Dialogue} in the summer of 2017. We
would like to thank Kevin Bassler, Ronald Dickman and Beate Schmittmann for helpful discussions.
\end{acknowledgements}

\bibliographystyle{spmpsci}      
\bibliography{WeissEtAl2019JSP}   

\end{document}